\documentclass{iopart}

\usepackage{amssymb}
\usepackage{graphicx}
\usepackage{psfrag,color}

\newcommand{\D}{\displaystyle}

\renewcommand\mark[1]{\bgroup\color{red}\bfseries{[#1]}\egroup}

\newcommand{\xx}[1]{#1}

\begin{document}

\title[]{Stress relaxation through crosslink unbinding in cytoskeletal networks}

\author{C. Heussinger} \address{Institute for Theoretical Physics,
  Georg-August-Universit\"at G\"ottingen, Friedrich-Hund-Platz 1, 37077
  G\"ottingen, Germany} \ead{claus.heussinger@theorie.physik.uni-goettingen.de}

\begin{abstract}
  The mechanical properties of cells are dominated by the cytoskeleton, an
  interconnected network of long elastic filaments. The connections between the
  filaments are provided by crosslinking proteins, which constitute, next to the
  filaments, the second important mechanical element of the network. An
  important aspect of cytoskeletal assemblies is their dynamic nature, which
  allows remodeling in response to external cues. The reversible nature of
  crosslink binding is an important mechanism that underlies these dynamical
  processes. Here, we develop a theoretical model that provides insight into how
  the mechanical properties of cytoskeletal networks may depend on their
  underlying constituting elements. We incorporate three important ingredients:
  nonaffine filament deformations in response to network strain; interplay
  between filament and crosslink mechanical properties; reversible crosslink
  (un)binding in response to imposed stress. With this we are able to
  self-consistently calculate the nonlinear modulus of the network as a function
  of deformation amplitude and crosslink as well as filament stiffnesses.
  During loading crosslink unbinding processes lead to a relaxation of stress
  and therefore to a reduction of the network modulus and eventually to network
  failure, when all crosslink are unbound. This softening due to crosslink
  unbinding generically competes with an inherent stiffening response, which may
  either be due to filament or crosslink nonlinear elasticity.

\end{abstract}

\submitto{\NJP}
\maketitle





\section{Introduction}

Cells display complex nonlinear and time-scale dependent rheological
properties~\cite{fletcher10:_cell,hoffman09:_cell_mechan,wen11:_polym}. A broad
range of relaxation timescales results in power-law spectra for the frequency
dependence of the linear viscoelastic response~\cite{KollmannsbergerReview2011}.
Under nonlinear loading conditions, cells can display apparently contradicting
behaviors, ranging from fluidization to
reinforcement~\cite{trepat2007Nature,Fernandez2006BPJFibroblast, WolffNJP2010}.
Understanding these mesoscale behaviors in terms of underlying non-equilibrium
processes, such as cytoskeletal remodeling, motor activity or reversible
crosslink binding or folding, remains an important theme in current
biomechanical research.

Over recent years, reconstituted f-actin networks have become a popular model
system in which these phenomena can be studied in detail. Much of previous
research in this field focused on the frequency-dependent rheology of
\emph{permanently crosslinked} filament
networks~\cite{bausch06,LielegSoftMatter2010,KaszaCOCB2007}. Key questions
revolve around the high-frequency modulus and its dependence on frequency
$\omega$~\cite{koenderinkPRL2006,gar04b}, the nature of network deformations
(affine vs.  non-affine) at intermediate freuqencies
\cite{PhysRevLett.98.198304,wen07:_local} and the nonlinear response properties
of the network~\cite{lielegPRL2007,gar04a}.
Theoretical models~\cite{mac95,gittesPRE1998,heussingerPRE2007} and simplified
simulation schemes~\cite{wil03,hea03a,Heussinger2006,onck05,PhysRevE.82.061902}
have been proposed that aim at explaining one or the other of these non-trivial
features.

In these studies, the filaments and their mechanical and thermal properties are
assumed to dominate the effective rheology of the system. This may be different
in F-actin networks crosslinked with the rather compliant crosslinking protein
filamin. Experimental and theoretical work
\cite{PhysRevLett.101.118103,PhysRevE.79.061914,kasza10:_actin_filam_lengt_tunes_elast}
suggest a second, crosslink-dominated regime, where network rheology is set by
the crosslink stiffness, while filaments effectively behave as rigid,
undeformable rods.
In between these two extreme scenarios a proper treatment would have to consider
the full interplay between crosslink and filament mechanical properties. This
has not been addressed theoretically before.

Recent experiments have indicated that at low driving frequencies effects due to
crosslink binding become important. This is evidenced, for example, as a peak in
the loss modulus $G''$~\cite{lielegPRL2008Transient} or a broad distribution of
time-scales leading to an anomalous scaling with frequency, $G''\sim
\omega^{1/2}$~\cite{broederszPRL2010Linker}. Theoretical modelling in this field
are only beginning to
emerge~\cite{lieleg11:_slow,WolffNJP2010,broederszPRL2010Linker}. Some of the
pertinent problems are: what is the force on a crosslink and how does it depend
on network deformation? how does this affect binding?  once unbound, how does
this affect the network ?

A noteworthy recent development is the phenomenological ``glassy wormlike
chain'' model~\cite{kroyNJP2007,Semmrich18122007}.  In that approach,
filament-filament interactions, as for example mediated by specific
crosslinking, are not modeled explicitly, but are assumed to lead to an
exponential stretching of the single filament relaxation times. Network
deformation is accounted for by a pre-stretching tensile force in the filament.
The same force is assumed to enter the unbinding rate constant of the crosslinks
via a Bell-like model~\cite{WolffNJP2010}.

Here, we go beyond previous studies by investigating the interplay between
filament and crosslink elasticity and its effects on the crosslink binding
behavior. We will present a simple model that accounts for the individual
crosslinks, their mechanical properties as well as their binding state. The
model is a fully thermodynamic treatment, where crosslink binding is
equilibrated for a given network deformation. No rate effects will be
considered. As a result of our analysis, we will be able to selfconsistently
calculate the nonlinear elastic modulus of the network, which incorporates as
key ingredient the ensuing tendency for crosslinks to unbind under load.

The manuscript is structured as follows. In section~\ref{sec:model} we will
define our model and relate it to existing approaches in the literature.  In the
model the filament network is represented as an effective elastic medium with a
given, fixed modulus $k_m$. We will introduce a Hamiltonian that describes the
properties of a test filament embedded in this medium. In
section~\ref{sec:results} we will present results of Metropolis Monte-Carlo
simulations for the Hamiltonian introduced. In section~\ref{sec:theor-fram} a
theoretical framework will be developed that allows some analytical results to
be obtained. Finally, in section~\ref{sec:selfc-determ-medi} we will discuss the
question of how to obtain in a self-consistent way the stiffness of the
effective medium in terms of the response properties of the test filament.

\section{Model} \label{sec:model}

We will consider the properties of a test filament crosslinked into a network.
The filament is described in terms of the worm-like chain model. In
``weakly-bending'' approximation the bending energy of the filament can be
written as
\begin{eqnarray}\label{eq:hb}
H_b &=&\frac{\kappa_b}{2}\int_0^L\left(\frac{\partial^2 y}{\partial
    s^2}\right)^2ds
\end{eqnarray}
where $\kappa_b$ is the filament bending stiffness and $y(s)$ is the transverse
deflection of the filament from its (straight) reference configuration at
$y_0(s)=0$. In these expressions $s$ is the arclength, $s=[0,L]$, and $L$ is the
length of the filament.

The effect of the surrounding network is to confine the test filament to a
tube-like region in space.  In this way the actual network is substituted by an
effective potential that acts on the test filament. A convenient potential is
the harmonic tube
\begin{eqnarray}\label{eq:tube}
V = \frac{1}{2}\int_0^L k(s) (y(s)-\bar y(s))^2 ds\,,
\end{eqnarray}
where $k(s)$ is the strength of the confinement and
$\bar y(s)$ is the tube center, which may or may not be different from
the reference configuration of the filament.

\begin{figure}[h]
\begin{center}
\includegraphics[clip=,width=0.9\columnwidth]{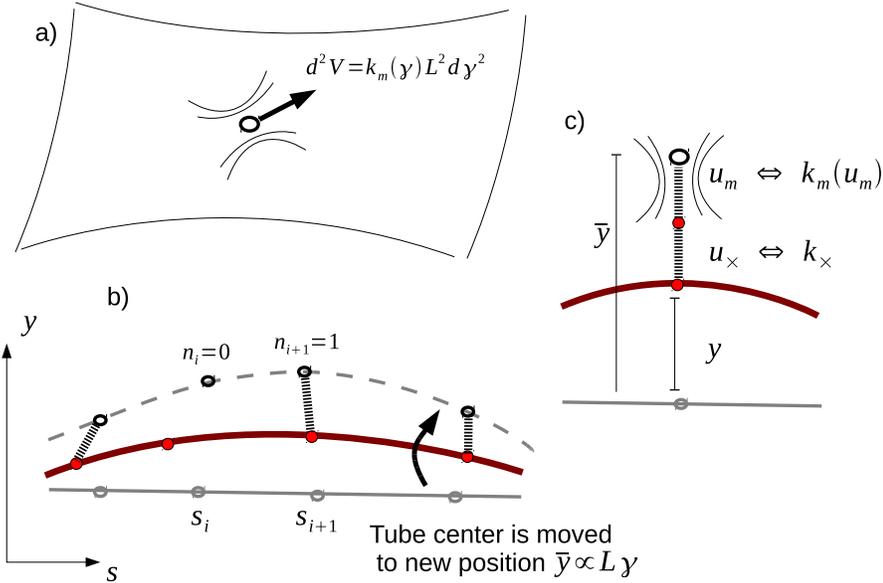}
\caption{Illustration of the modeling assumptions. a) On a macroscopic level the
  filament network is modeled as a (nonlinear) elastic medium with modulus
  $k_m(\gamma)$. b) On the microscopic level the effect of the network is to
  confine the filament to a tube-like region in space. In a densly crosslinked
  network the tube potential mainly acts at the crosslinking sites $i=1\ldots
  N_\times$, which are located at discrete points $s_i$ along the filament axis
  and which may be bound ($n_i=1$) or unbound ($n_i=0$). A macroscopic strain
  $\gamma$ leads to an inhomogeneous distortion of the effective medium, which
  is accounted for by a shift of the tube center line, $\bar y \propto L\gamma$
  (dashed line). The filament length $L$ plays the role of a nonaffinity
  length-scale, i.e. it specifies the length-scale at which medium deformations
  are inhomogeneous. The filament (with its bending stiffness $\kappa_b$)
  resists the tube deformation, and leads to a frustration effect between
  filament bending and crosslink deformation. With the possibility of crosslink
  unbinding ($n_i=1 \to 0$) this frustration is avoided.  c) The tube potential
  itself consists of a contribution from the crosslinks (with stiffness
  $k_\times$) and from the effective medium $k_m$. The total deformation $\bar
  y$ is shared between the three elements filament, crosslink and medium, $\bar
  y = y + u_\times + u_m$.
\label{illustration}}
\end{center}
\end{figure}

The tube potential is a convenient representation of many-body network effects
on the test filament. However, the parameters entering the potential (here
$k$ and $\bar y$) and their relation to network deformation and
mechanical properties are usually unknown.  Ideally, these parameters can be
determined in a self-consistent way from the analysis of the test filament.

In the following, we will present a theoretical framework, where such a
self-consistent determination is partly possible (see Fig.~\ref{illustration}
for a short description). In this model we assume the network to be represented
by an effective medium that couples to the test filament only at the
crosslinking points,
\begin{eqnarray}\label{eq:kx}
k(s) = k_\times\sum_{i=1}^{N_\times}n_i \delta(s-s_i)
\end{eqnarray}
where $k_\times$ is the crosslink stiffness. $N_\times$ is the total number of
crosslinking sites. Being interested in the effects of reversible crosslink
binding, we also include an occupation variable $n_i=0,1$. If the crosslink is
bound at site $i$, we have $n_i=1$, if it is unbound then $n_i=0$. Below, we
will also introduce (and calculate self-consistently) the modulus $k_m$ of the
effective medium. The crosslink stiffness $k_\times$ in equation~(\ref{eq:kx})
then has to be substituted by an effective stiffness $k_{\rm
  eff}(k_\times,k_m)$, which contains contributions from both, the crosslinks
and the effective medium. For the time being we set $k_m\to\infty$ and thus
$k_{\rm eff}\to k_\times$.

As we want to discuss the role of network deformation on the test filament, we
need to know how a macroscopically applied strain field couples to the single
filaments. If the effective medium can be thought of as strictly homogeneously elastic then
the local strain felt by the filaments would be identical to the macroscopic
strain $\gamma$ imposed by the experimental device. With this assumption of
``affine'' deformations any network deformation couples to the end-to-end
distance of the filament and leads to its extension or compression. Rheological
properties are then governed by the resistance of filaments to changes in their
end-to-end distance\cite{mac95,Storm2005,Gardel2004,PhysRevE.79.061914}.

An alternative point of view is based on the fact, that filaments are
anisotropic elastic objects~\cite{Kroy1996}, where bending is a much softer
deformation mode than stretching~\footnote{With an effective bending stiffness
  $k_b\sim \kappa_b/l^3$, and a stretching stiffness $k_s\sim \kappa_bl_p/l^4$
  we find $k_b/k_s\sim l/l_p<1$ as the persistence length $l_p$ is usually much
  larger than the relevant filament length $l$. }. This, together with local
structural heterogeneity, may in fact lead to a highly nonaffine response where
network properties are determined by the resistance of filaments to bending
deformations
\cite{heu06floppy,heussingerPRE2007,heussingerEPJE2007,lielegPRL2007,onck05}.

It has been pointed out in~\cite{heussingerPRE2007} that the filament length $L$
may play the role of a lower cut-off at which the affine assumption breaks down.
\xx{This breakdown is due to inextensibility and geometric correlations that
  develop along essentially straight filaments.  The actual deformations on the
  smaller scale} of the crosslink (or the mesh-size) then follow from the local
network structure.  Adopting this latter point of view we assume that
macroscopic and homogeneous network deformations (as measured by a strain
$\gamma$) locally lead to an inhomogeneous distortion of the effective medium
that can be accounted for by a strain-dependent tube center line $\bar
y(s)\propto L\gamma$. The occurence of the filament length $L$ in this equation
is a direct consequence of its role as a nonaffinity length-scale.  \xx{For our
  purposes the scaling with $L$ is not decisive, however. What is more important
  here is that nonaffine deformations of the tube increase with network strain
  $\gamma$ and are slaved to the local network structure.}

In terms of our discrete crosslinking points this means that one head of the
crosslink follows the effective medium deformation at $\bar y_i$, while the
second head remains on the filament at $y_i$.  Whenever the filament does not
follow the medium, it is the crosslinks that have to deform by $u_\times=\bar
y_i - y_i$ leading to an elastic energy $~k_\times (\bar y_i - y_i)^2$. The tube
potential is therefore identified with the crosslink deformation energy.

The model thus describes an interplay between crosslink deformation and filament
bending. A macroscopic strain applied to the background medium leads to a
frustration effect between crosslink and filament deformations.  With the
possibility of crosslink unbinding, this frustration is avoided. The price to
pay is the binding enthalphy $H_{\rm bind}= -|\mu| \sum_i n_i$, which is assumed
to favour the bound crosslink state with $n_i=1$.

A similar competition between filament and crosslink energies is present in
ordered arrays of parallel filaments or filament
bundles~\cite{PhysRevE.83.050902}.  In that case, the role of the medium strain
is taken by a deformation of the bundle as a whole. This deformation induces
filament sliding and leads to a frustration effect between filament stretch and
crosslink shear. Subsequently, and to relieve this frustration, a discontinuous
crosslink unbinding transition is observed.  Interestingly, this has the form of
an unzipping transition~\cite{vink12:_cross}.  Crosslinks unbind first at places
where bundle deformation is largest. An interface then moves rapidly to the
other end of the bundle.


In the following we will analyze the model defined in equations (\ref{eq:hb}) to
(\ref{eq:kx}) with the aim of understanding the combined effects of filament
deformation and crosslink binding in response to a network strain. We will
discuss some thermodynamic parameters as a function of strain, in particular the
average crosslink occupation, $\langle n\rangle$ and the filament energy $E$.
The latter will provide the connection to the stiffness of the network and can
be determined in rheological experiments. 


\section{Monte-Carlo simulations} \label{sec:results}

For the Monte-Carlo simulations we represent the filament by a one-dimensional
lattice.  To each lattice site $(i=1,\ldots,N_\times)$ the pair $(y_{i},n_i)$ is
attached, denoting the local transverse displacement $y_i$ and the crosslink
occupation variables $n_i$.

As Monte Carlo moves we use single site displacements and crosslink
binding/unbinding moves, the latter attempted with ten percent probability. In a
displacement move, a lattice site $i$ is selected randomly, and the current
displacement $y_i$ of that site is replaced by $y_i' = y_i + \delta$, with
$-0.1a < \delta < 0.1a$ drawn uniformly randomly, and $a=L/N_\times$ is the
discretization length, which serves as unit of length here. The new displacement
$y_i'$ is accepted with probability $P_{\rm disp} = \min \left[1, e^{-\beta
    \Delta H} \right]$, where $\Delta H$ is the energy difference between
initial and final state. During a crosslink move, a bond is selected randomly,
and the corresponding occupation variable $n_i$ is flipped ($n_i\to 1-n_i$). The
new state is accepted with probability $P_{\rm xlink} = \min \left[1, e^{-\beta
    \Delta H + \beta \mu \Delta N} \right]$, with $\mu$ the crosslink chemical
potential, $\Delta N = \pm 1$ the change in the number of crosslinks.

In the following we show data where the filament bending stiffness $\kappa_b=1$
serves as unit of energy. The chemical potential is $|\mu|=0.001$. The inverse
temperature is $\beta=1000$ such that the persistence length is
$l_p=\beta\kappa=1000$, measured in units of lattice sites $a$. The filament
length is taken to be $L=20a$, i.e. $N_\times=20$. For simplicity, we assume the
filament to have zero deflection at its ends, $y_0=y_{N_\times}=0$.  The tube
potential is taken as $\bar y(s)=\gamma L \sin(qs)$ with $q=\pi/L$, which
corresponds to the longest possible wavelength compatible with the chosen
boundary conditions.

\begin{figure}[h]
\begin{center}
\includegraphics[clip=,width=0.49\columnwidth]{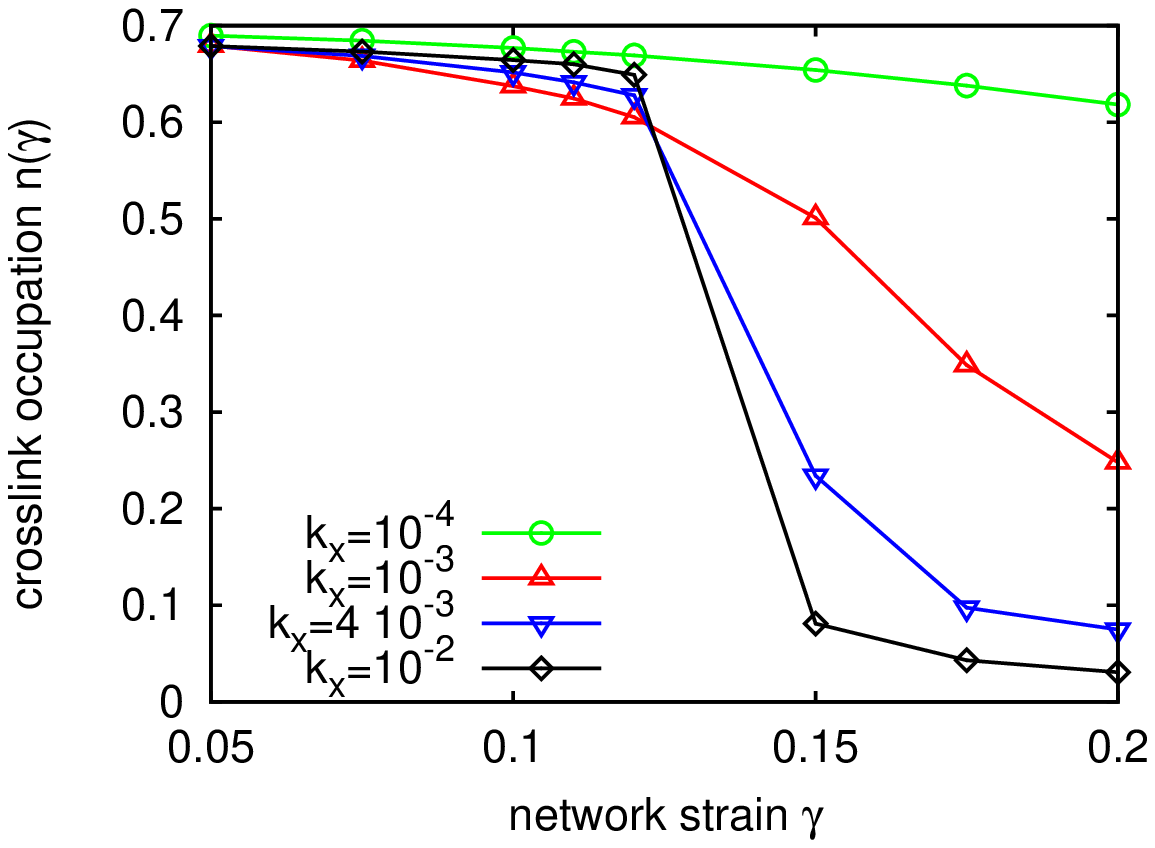}
\includegraphics[clip=,width=0.49\columnwidth]{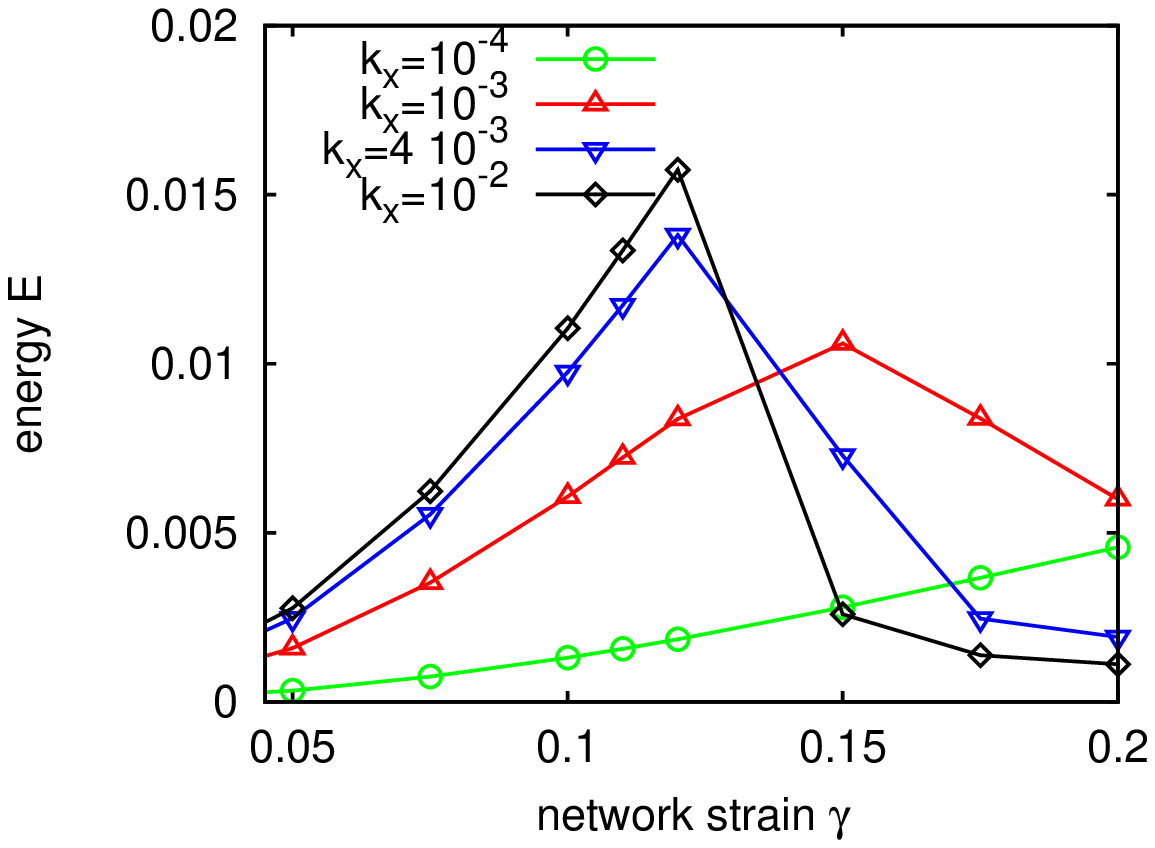}
\caption{Average crosslink occupation $n$ (left) and elastic energy $E$ (right)
  as a function of network strain $\gamma$ and for different crosslink stiffness
  $k_\times$. Stiff crosslinks unbind in a sudden discontinuous transition. The
  elastic energy stored in the filament is lost.\label{n.energy}}
\end{center}
\end{figure}

In Fig.\ref{n.energy} we monitor the average crosslink occupation $n =\sum_i
\langle n_i\rangle /N_\times$ as well as an average energy $E$, which is
obtained by minimizing the total elastic energy for given crosslink occupation.
We vary the network strain $\gamma$ as well as the crosslink stiffness
$k_\times$. It can readily be seen that crosslink stiffness has a dramatic
effect on the thermodynamic state of the system.

For mechanically weak crosslinks network strain $\gamma$ has no strong influence
on the binding state. Only few crosslinks unbind upon increasing network
deformation. In this regime the crosslinks are not strong enough to enforce
filament bending.  The elastic energy is small, and primarily stored in the
crosslinks. As a consequence, the energy of the bound state is raised and the
statistical weight is shifted towards the unbound state.  This is the regime
discussed in
~\cite{PhysRevLett.101.118103,PhysRevE.79.061914,kasza10:_actin_filam_lengt_tunes_elast}
in the context of filamin-crosslinked f-actin networks. 
Note, that in that model no crosslink unbinding is accounted for. The crosslinks
are modeled as nonlinear elastic elements, leading to significant stiffening of
the network under strain. We have checked that incorporating nonlinear crosslink
compliance in our model reproduces this behavior. Moreover, with the possibility
of crosslink unbinding, this stiffening due to crosslink mechanics generally
competes with a softening effect due to crosslink unbinding.

When the crosslinks are sufficiently strong, their deformation energy starts to
compete with the filament bending energy. Now crosslinks can force filaments
into deformation and the elastic energy is mainly stored in the filaments. At
large network strains this energy is too high, however, and unbinding becomes
favourable.  This is evident as a discontinuous unbinding transition, where
nearly all remaining crosslinks unbind simultaneously. Associated with such a
transition is a free-energy barrier. The escape time over this barrier sets the
time-scale of relaxation of the imposed deformation mode.  This will be
exponential in the number of crosslinking sites $N_\times$, reminiscent of what
has been proposed in \cite{kroyNJP2007}.

\section{Theoretical framework}\label{sec:theor-fram}

Theoretical progress can be made in the continuum limit of equations
(\ref{eq:hb}) to (\ref{eq:kx}).  Making the mean-field assumption $n_i\to \sum_i
n_i/N_\times\equiv n$, the energy of the test filament can be written as
\begin{equation}\label{eq:hamiltonian.continuum}
H = \frac{\kappa_b}{2}\int y''(s)^2 ds  +
    \frac{k_\times n}{2a}\int \left(y(s)-\bar y(s)\right)^2 ds\,.
\end{equation}

Going to Fourier space, $y(s)=\sum_i y_q\sin(qs)$, the $y_q$ can be integrated
out. The resulting effective free energy $F(n)$ can be written as
\begin{eqnarray}\label{eq:free.energy}
e^{-\beta F(n)} = \left(\prod_q\frac{4\pi/\beta L}{\kappa_bq^4+k_\times
    n/a}\right)^{1/2}\exp(-\beta N_\times(F_\gamma(n)+\mu n))
\end{eqnarray}
with a deformation-dependent part $F_\gamma(n)$
\begin{eqnarray}\label{eq:free.energy.A}
F_\gamma(n) =\sum_q\frac{A_q(\gamma)}{1+n_q/n}
\end{eqnarray}
a deformation amplitude $A_q(\gamma)=(a/4)\kappa_bq^4\bar y_q(\gamma)^2$ and
the scale for crosslink density $n_q=\kappa_bq^4a/k_\times$.

The prefactor $(..)^{1/2}$ is an entropic contribution that specifies the entropic cost of
binding to crosslinks. Clearly, binding suppresses bending undulations and
therefore reduces the entropy stored in these modes. The interplay between
entropy reduction and $\mu$-dependent enthalpy gain may lead to a thermal
unbinding transition~\cite{kierfeld05PRL,benetatosPRE2003}. For the case
considered here such an unbinding transition is not relevant. Instead, we want
to focus on the deformation-dependent part $F_\gamma(n)$.

Just as in the simulations, we make the single-mode assumption,
$A_q(\gamma)=A\delta_{qq_1}$ and $q_1=\pi/L$. Then it is easy to see that the
free energy has two saddle-points at $n=0$ and $n=1$ and a discontinuous
transition between them at $A^\star=\mu(1+n_{q_1})$. This condition gives a
critical network strain of $\gamma^\star=0.13$, which compares well with the
actual transition as seen in figure~\ref{n.energy}. This analysis predicts a
discontinuous unbinding transition for any value of crosslink stiffness
$k_\times$, in apparent disagreement with the simulations.  It turns out that
this is a result of the saddle-point approximation. In a more refined treatment,
which also includes an ``entropy of mixing'' term, $p_n=\left({N_\times \atop
    nN_\times}\right)$, the discontinuous transition is destroyed when the
crosslinks are sufficiently soft~\cite{PhysRevE.83.050902}.

Finally, let us comment on the use of the single-mode assumption for the tube
center $\bar y(s)=\sum_i\bar y_q(\gamma)\sin(qs)$. The precise form of $\bar
y(s)$ and its dependence on network strain $\gamma$ is, in principle, unkown and
one would like to calculate it selfconsistently. As this is unfeasible, one has
to rely on assumptions and physical plausibility. Conceptually, the tube
deformation in response to network strain $\gamma$ represents the missing link
between affine deformations on scales larger than the filament length and the
actual crosslink motion on the scale of the mesh-size. As such it will be
sensitive to the local structure of the network. In \cite{heussingerPRE2007} it
has been shown how such a link can be constructed in terms of local binding
angles and the mesh-size distribution.

For our purposes we note, that using one or few higher modes will not
fundamentally alter the proposed picture of continuous vs.  discontinuous
crosslink unbinding.  The necessary condition for such a feature to persist is a
free energy contribution $F_\gamma(n)$ that grows slower than linear with
crosslink occupation $n$.  In this case the linear contribution $-|\mu|n$ from
the binding enthalpy will eventually take over.

With one or a finite number of modes present the free energy will eventually
saturate $F_\gamma(n)= \rm const$. This happens, when the filament is strongly
constrained by the crosslinks to precisely follow the tube centerline $y(s)=\bar
y(s)$. Clearly, binding even more crosslinks cannot lead to a more efficient
confinement.
A similar result is obtained, if one assumes infinitely many modes, with an
amplitude that depends on mode-number as $\bar y_q^2\sim q^{-4}$, i.e. like a
thermalized bending mode of a worm-like chain. In this case, the free energy
does not saturate but asymptotically grows like $F_\gamma(n)\sim n^{1/4}$, i.e.
also slower than linear.




\section{Selfconsistent determination of medium stiffness}\label{sec:selfc-determ-medi}

In the previous sections we assumed the test filament to be coupled to a medium
of infinite stiffness. In reality the medium itself is made of crosslinked
filaments. Thus the medium properties cannot be set externally, but should be
determined self-consistently from the properties of the test filament and its
crosslinks. In particular, the studied crosslink unbinding processes reduce the
connectivity, and therefore stiffness, of the medium. Unbinding should,
therefore, be reflected as a change in the tube potential.


In the following, we will therefore assume the medium to be characterized by an
energy function $V_m(u_m)$, which quantifies the energy cost of a deformation
$u_m$. The total tube potential $V=V_m +V_\times$ then contains contributions
from both the crosslinks \emph{and} the effective medium. As we expect the
medium to be nonlinear, $V_m$ is not necessarily a harmonic function of
deformation amplitude.


With a finite medium compliance, any transverse displacements, $\bar y_i-y_i$ at
a crosslink, has to be shared between the crosslink and the medium, $\bar
y_i-y_i = u_\times + u_m$.  The relative stiffness of the two elements dictate
the magnitude of the deformations via a force balance
condition~\cite{PhysRevE.79.061914}.
With this the total confinement strength is no longer set by the crosslink
stiffness, $k_\times$, but by an effective stiffness $k_{\rm eff}$ determined by
the serial connection of the crosslink and the medium. Unbinding events are
expected to reduce the stiffness of the medium and therefore of $k_{\rm eff}$.
In a softer environment, however, the test filament and its crosslinks will have
a \emph{reduced} tendency for further crosslink unbinding. There is, therefore,
a negative feedback loop between medium stiffness and crosslink unbinding. This
may smooth out the sudden unbinding transition that was observed in figure
\ref{n.energy}.


To calculate $V_m$ we need a condition of self-consistency. This is based on
previous approaches \cite{heussingerPRE2007,PhysRevE.79.061914}.  As discussed
above, a macroscopic strain $\gamma$ leads to a local medium deformation of
$\bar y=\gamma L$. The associated energy cost is given by $V_m(\gamma L)$. At
the same time the strain deforms the test filament and its surrounding tube.
Therefore, the energy that is needed to impose the strain $\gamma$ needs to be
balanced by the energy $E$ that is build up in the test filament.  This
condition can be written in the simple form
\begin{eqnarray}\label{eq:km.sc} V_m(\gamma L) = E
\end{eqnarray}
where $E=\langle H_b+V_\times + V_m\rangle_n$ and the average is taken over
crosslink occupation $n$ and network disorder. Despite its simple form,
equation~(\ref{eq:km.sc}) is rather difficult to handle, as the unknown
potential $V_m$ is required for the evaluation of the statistical average. A
possible solution could proceed iteratively, starting from a suitably chosen
initial guess. We have not tried such a scheme. Instead, and to make analytical
progress, we propose a simplified Ansatz for the confinement potential. As we
will see, this Ansatz successfully describes the intuitive result of a softening
of the medium as a result of crosslink unbinding. 

Let us assume the confinement to be harmonic in the $y$-degrees of
freedom
\begin{eqnarray}\label{eq:v.nonlinear.harmonic}
V = \frac{1}{2}\sum n_i k_{\rm eff} (y_i-\bar y_i)^2
\end{eqnarray}
but with a confinement strength
\begin{eqnarray}\label{eq:serial_springs}
\frac{1}{k_{\rm eff}} = \frac{1}{k_\times} +\frac{1}{k_m(\gamma)}
\end{eqnarray}
that depends on the deformation amplitude $\gamma$. The form of equation
(\ref{eq:serial_springs}) mimics the serial connection of crosslink and medium.

The energy $E$ can then easily be calculated, and equation~(\ref{eq:km.sc}) takes
the simple form
\begin{eqnarray}\label{eq:km.sc.2}
  k_m=\left\langle L\left(\frac{1}{\kappa q^4}+\frac{a}{nk_{\rm eff}}   \right)^{-1}\right\rangle_n\,,
\end{eqnarray}
which has to be solved for $k_m(\gamma)$. The dependence on strain $\gamma$ is
implicit in the averaging procedure, as the tube potential
(\ref{eq:v.nonlinear.harmonic}) depends on $\gamma$.
Thus, the effective medium stiffness is obtained as a serial connection of the
three mechanical elements fiber, $k_\perp\sim \kappa_b/L^3$, crosslinks,
$nk_\times$, and the medium itself, $nk_m$.

\begin{figure}[h]
\begin{center}
\includegraphics[clip=,width=0.5\columnwidth]{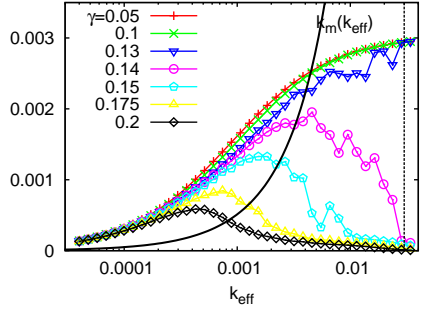}
\caption{Graphical solution to equation (\ref{eq:km.sc.2}) for $k_\times =
  0.04$. The right-hand side of the equation (symbols) is drawn for different
  deformation amplitudes $\gamma$. The network modulus $k_m$ is found by
  intersecting with the left-hand side of the equation (solid line), $k_m(k_{\rm
    eff})$. The vertical dashed line corresponds to the solution without the
  self-consistency condition, i.e. with a prescribed medium stiffness, as
  discussed in section~\ref{sec:results}.\label{graphical.sol}}
\end{center}
\end{figure}

Figure \ref{graphical.sol} presents a graphical solution of equation
(\ref{eq:km.sc.2}). The symbols give the effective fiber stiffness (right-hand
side of the equation, from the MC simulations) plotted as a function of $k_{\rm
  eff}$ and taken at different deformation amplitudes $\gamma$.  If we assume
$k_{\rm eff}$ to be given and constant (as in figure \ref{n.energy}), a vertical
line would give us the fiber stiffness as a function of amplitude $\gamma$. For
$k_{\rm eff}$ large enough (indicated by the vertical dashed line), a
discontinuous transition is evident in the data. To extract the actual network
modulus $k_m(\gamma)$ we have to find the intersection with the curve
$k_m(k_{\rm eff})$, equation (\ref{eq:serial_springs}), which is drawn as solid
black line. The resulting $k_m$ is clearly decreasing with deformation
amplitude, however, the discontinuous nature is not obvious anymore.

\begin{figure}[h]
\begin{center}
\includegraphics[clip=,width=0.49\columnwidth]{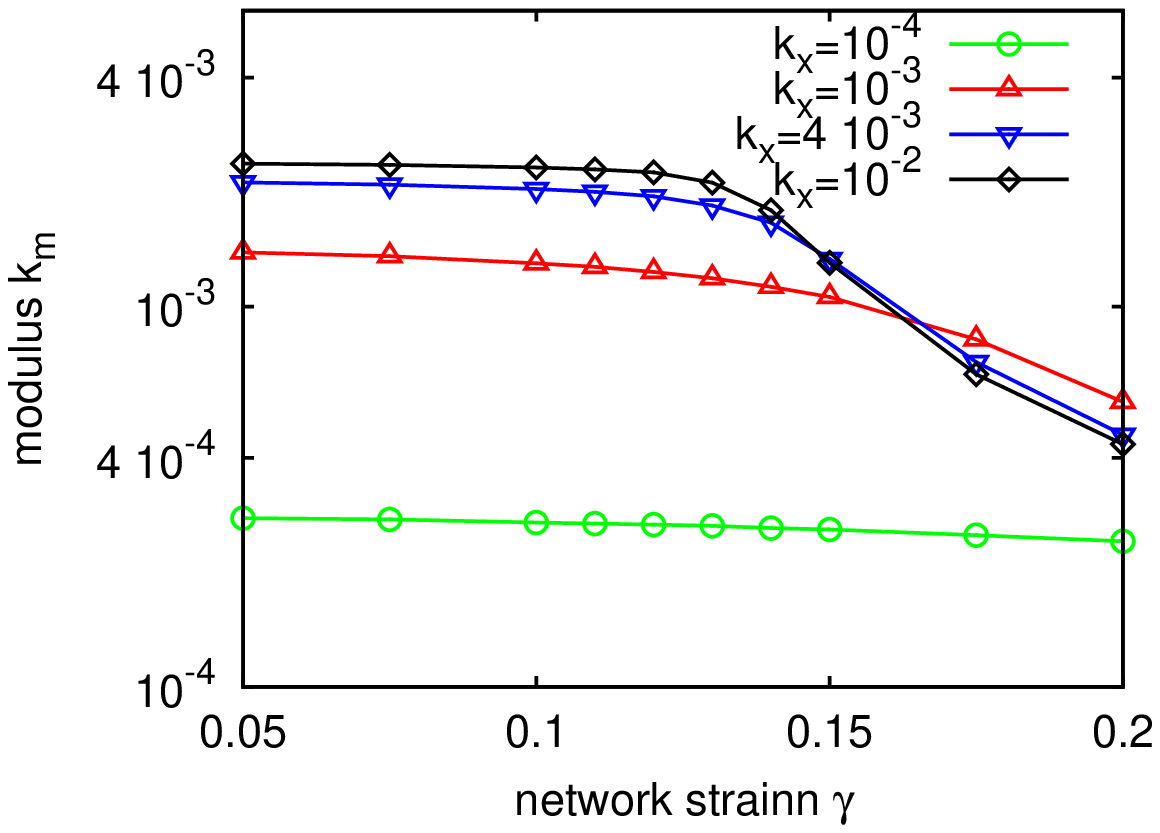}
\includegraphics[clip=,width=0.49\columnwidth]{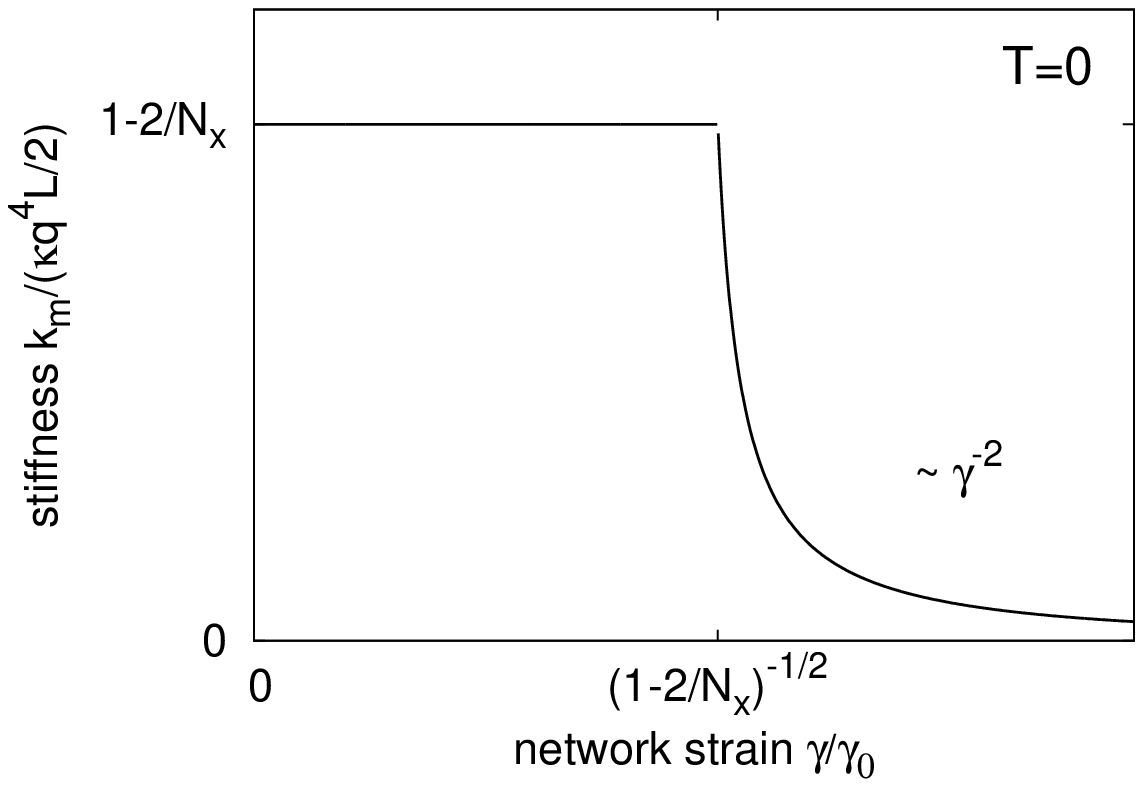}
\caption{(left) Network modulus $k_m(\gamma)$ for different crosslink stiffness
  $k_\times$ as obtained from the MC simulation. (right) Analytical result for
  $k_\times=\infty$ and based on the zero temperature approximation presented in
  section~\ref{sec:theor-fram}. \label{km.sim.th}}
\end{center}
\end{figure}

A quantitative analysis is presented in figure~\ref{km.sim.th}. The resulting
network modulus $k_m(\gamma)$ is plotted for different crosslink stiffness
$k_\times$. For small strain the modulus $k_m$ is constant, as for a linear
elastic material.  This linear stiffness first increases with crosslink
stiffness but then saturates, when the crosslinks become stiffer than the fiber.
This behavior is in line with the serial connection between fiber, crosslink and
medium as embodied in equation~(\ref{eq:km.sc.2}). In a serial connection it is
always the softer element that governs the mechanical properties.

At higher strain and for stiff crosslinks, the network modulus $k_m$ decreases.
This decrease is not discontinuous as expected from figure \ref{n.energy}, but
rather smooth and gradual.

In fact, a zero-temperature analysis along the lines of
section~\ref{sec:theor-fram} shows that the discontinuity turns into a
second-order transition, with a cusp as indicated in the right panel of figure
\ref{km.sim.th}. In the limit, $k_\times\to\infty$, we find
\begin{eqnarray}\label{eq:km_theory}
  k_m(\gamma) = \frac{\kappa q^4L}{2}\cdot
\cases{(1-2/N_\times)&for $\gamma<\gamma^\star$\\
\frac{\D 2/N_\times}{\D(\gamma/\gamma_0)^2-1} & for $\gamma\ge \gamma^\star$\\}
\end{eqnarray}
with $\gamma_0^2=4\mu/\kappa q^4L^2a$ and the critical amplitude
$\gamma^\star=\gamma_0(1-2/N_\times)^{-1/2}$.

\begin{figure}[h]
\begin{center}
\includegraphics[clip=,width=0.49\columnwidth]{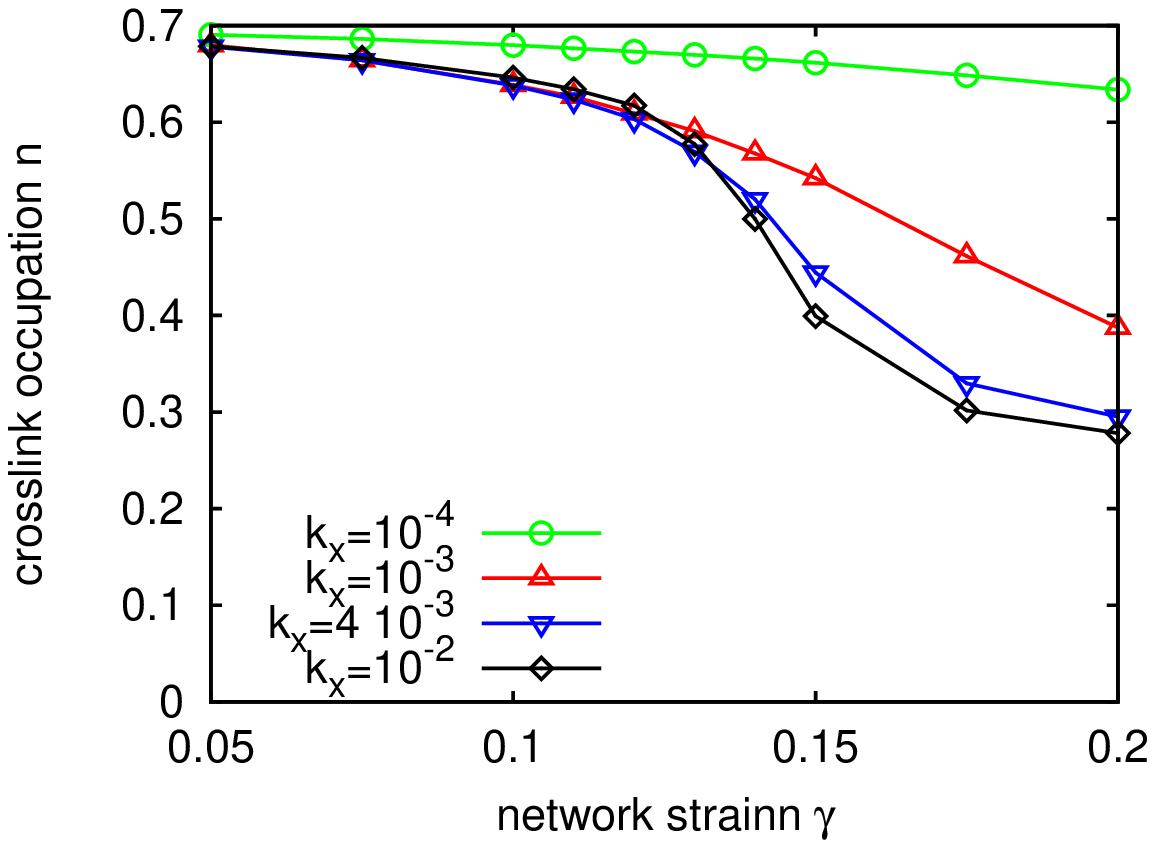}
\includegraphics[clip=,width=0.49\columnwidth]{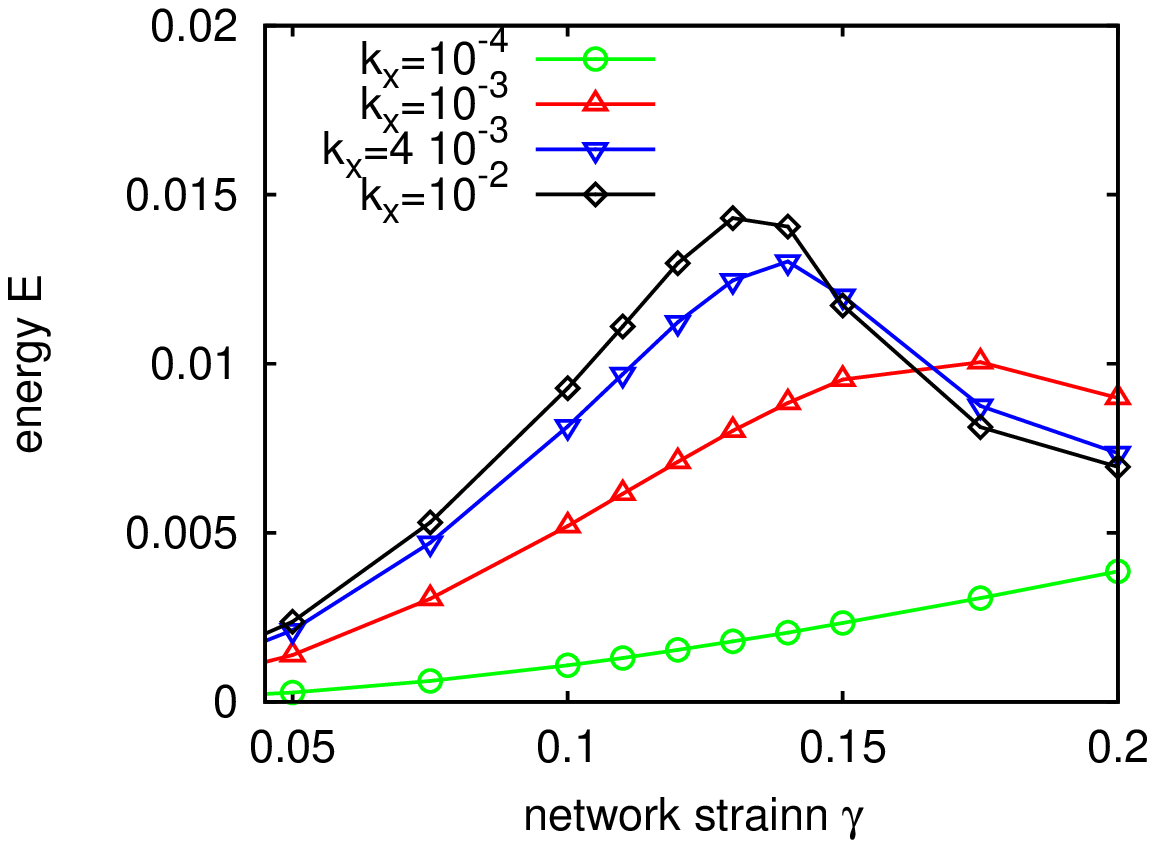}
\caption{Average crosslink occupation $n$ (left) and elastic energy $E$ (right)
  as a function of network strain $\gamma$ and for various crosslink stiffness
  $k_\times$. Stiff crosslinks tend to unbind under strain leading to an associated
  reduction in elastic energy. As compared to figure~\ref{n.energy}, however,
  unbinding is not dramatically discontinuous but rather smooth.
  \label{n.energy.sc}}
\end{center}
\end{figure}

Finally, figure \ref{n.energy.sc} shows the resulting crosslink occupation $n$
as well as the average energy of the test filament.  Qualitatively, the
conclusion is similar as in figure \ref{n.energy}. The crosslink stiffness is
identified as key factor in mediating crosslink unbinding processes.
Quantitatively, we see that the unbinding under load is more gradual in this
second scenario, where the medium stiffness is determined self-consistently.
This reflects the anticipated negative feedback of medium stiffness on crosslink
unbinding. 

Looking back at figure \ref{graphical.sol} it is clear that the strength of the
variation in $k_m$ or $n$ depends on the relative location of the two curves,
$k_m(k_{\rm eff})$ and the effective fiber stiffness as embodied in the
right-hand side of equation~(\ref{eq:km.sc.2}). For this geometrical factors
could be important. These could arise from network structural randomness, like
bond angles or crosslink distances. In our calculation these geometrical factors
have been disregarded, which corresponds to a regular network architecture.

\section{Conclusion}\label{sec:conclusion}

In this study, we discussed the interplay between filament and crosslink
elasticity in semiflexible polymer networks. In particular, we were interested
in the force-induced unbinding of crosslinks in response to external load.
Importantly, we considered the limiting case of slowly changing load
(``quasistatic''), where the system is given time enough to reach an equilibrium
state at each load level. The model presented is therefore purely thermodynamic
in nature, and no rate constants for the crosslink dynamics are needed.
Possible extensions of the present work may include the effect of a load applied
at finite rates.


We model the filament network as an elastic medium with modulus $k_m$.  The
stiffness is calculated on a self-consistent basis from the response of a test
filament that is embedded into the medium. On the microscopic level the effect
of the network is to confine the filament to a tube-like region in space.



We quantify the applied load in terms of a network strain $\gamma$, which is
homogeneous on a macroscopic scale. On the local scale of the individual
filaments, however, even a homogeneous strain leads to an inhomogeneous
(``nonaffine'') deformation of the effective medium. This is a natural
consequence of network heterogeneity and filament mechanical anisotropy. We
modeled this inhomogeneous deformation in terms of a distortion of the
center-line of the confinement tube of the filaments.
The filament (with its bending stiffness $\kappa_b$) resists this tube
deformation, which leads to a frustration effect between filament bending,
crosslink deformation and medium deformation. 

This competition is formalized in equation (\ref{eq:km.sc.2}) which we rewrite
here as
\begin{eqnarray}\label{eq:km.sc.2.appendix}
  k_m=\left\langle\left[\frac{1}{k_\perp}+\frac{1}{Nk_\times}+\frac{1}{Nk_m}  \right]^{-1}\right\rangle_N\,.
\end{eqnarray}
with an effective filament bending stiffness $k_\perp\sim \kappa_b/L^3$, the
crosslink stiffness $k_\times$ and the network modulus $k_m$. By solving this
equation, the latter is thereby obtained self-consistently from a serial
connection of the filament, the crosslinks and the medium itself. The most
interesting feature in this equation is the dependence on the number of bound
crosslinks $N$.  With the possibility of crosslink unbinding (decreasing $N$)
the competition between the different mechanical elements is avoided. This leads
to a reduction of the medium stiffness with increasing network strain, and
eventually to network failure, when all crosslinks are unbound.

Different scenarios can be distinguished. If the crosslinks are soft ($k_\times$
small), then the network modulus is dominated by the crosslinks, $k_m\sim
k_\times$, and the filament effectively behaves as a rigid rod. The tendency for
crosslink unbinding is weak as it does not lead to a significant stress
relaxation.
%
%
On the other hand, if crosslink and filament stiffness compete ($k_\times$
large), then unbinding events do help relax the imposed stress and reduce
the amount of stored energy.  This unbinding can be sudden and discontinuous or
take the form of a second-order transition, where $k_m(\gamma)$ or $N(\gamma)$
display a kink at a critical load $\gamma^\star$.

\xx{Experimentally, strain-induced stress relaxation is observed in living cells
  after transient pulses of stretch~\cite{trepat2007Nature} or in-vitro when the
  loading rates are small~\cite{LielegSoftMatter2010}. For larger loading rates,
  a pronounced stiffening is found in the in-vitro system. This is consistent
  with the first-order crosslink unbinding scenario discussed here.  In this
  picture, a free-energy barrier, and the associated time-scale, prevents
  crosslink unbinding when the loading rate is too large. Related phenomena are
  important for the aging behavior of kinetically trapped actin networks
  \cite{lieleg11:_slow}, where built in stresses only relax slowly and by the
  action of crosslink binding events. Similarly, red blood cells owe their
  remarkable ability to undergo reversible shape changes to a rewiring of the
  spectrin network~\cite{gov07:_activ,li07:_cytos}.  }

This work goes beyond previous models in considering both the filament and the
crosslink stiffness as factors for the rheological properties of crosslinked
filament networks. Moreover, we show how this interplay affects the tendency of
crosslinks to bind/unbind from the filaments during a rheological experiment.
The strain field imposed by the rheometer leads to nonaffine deformations on the
scale of the filaments. As compared to previous models, the unrealistic
assumption of affine deformations is abandoned in favour of a model that
incorporates the filament length as the fundamental non-affinity scale.



In extensions of the present model one should incorporate nonlinear elastic
compliances of the filaments and of the crosslinks. This is believed to be
important for the nonlinear strain stiffening of f-actin networks. The complex
interplay between stiffening and softening described in rheological
experiments~\cite{LielegSoftMatter2010} would then reflect the relative
importance of filament/crosslink elasticity, which lead to stiffening, and
softening as due to crosslink unbinding. \xx{Interestingly, unbinding processes
  may under some conditions also lead to the reverse (i.e. stiffening)
  effect~\cite{gov07:_less}. This further contributes to the rich nonlinear
  behavior of these systems, which is far from being fully understood.}

\ack

Support by the Deutsche Forschungsgemeinschaft, Emmy Noether program: He
6322/1-1, and by the collaborative research center SFB 937 is acknowledged.


\end{document}